# High sensitivity magnetic imaging using an array of spins in diamond

3. Physikalisches Institut, Universität Stuttgart, Pfaffenwaldring 57, 70550 Stuttgart

Steinert S.[1], Dolde F., Neumann P., Aird A., Naydenov B., Balasubramanian G., Jelezko F. and Wrachtrup J.

[1]To whom correspondence should be addressed: Email: s.steinert@physik.uni-stuttgart.de

**Abstract:**

We present a solid state magnetic field imaging technique using a two dimensional array of spins in diamond. The magnetic sensing spin array is made of nitrogen-vacancy (NV) centers created at shallow depths. Their optical response is used for measuring external magnetic fields in close proximity. Optically detected magnetic resonance (ODMR) is readout from a 60x60 μm field of view in a multiplexed manner using a CCD camera. We experimentally demonstrate full two-dimensional vector imaging of the magnetic field produced by a pair of current carrying micro-wires. The presented widefield NV magnetometer offers in addition to its high magnetic sensitivity and vector reconstruction, an unprecedented spatio-temporal resolution and functionality at room temperature.

**I. Introduction**

Sensing and imaging magnetic fields has been developed extensively during the last few decades propelled by applications ranging from medical to materials science. A wealth of methods has been developed to sense and image magnetic fields with high resolution and sensitivity. Magnetic resonance imaging (MRI), an inevitable tool in clinical diagnostics, would have revolutionary impact if the sensitivity could be improved to image live cells with sub-cellular resolutions. The inherent limitation arises from the conventional inductive detection that results in a poor signal to noise ratio.

Alternative detection techniques using superconducting quantum interference devices[1], magnetic resonance force microscopy[2-3] and alkali vapor atomic magnetometers[4-5] have been explored. Though these methods offer high sensitivity, they require cryogenic cooling, vacuum conditions or prolonged data acquisition and are therefore limited to special applications.

Alternatively, a proximal single nitrogen vacancy center (NV) in diamond as magnetic sensing probe showed promising results.[6-10] The NV defect center has several outstanding advantages for this application. Optical polarization and spin state readout facilitates convenient and non-invasive far field detection[11]. Furthermore, the atomic NV sensor permits the minimization of the sample-sensor distance, while the long spin coherence in the nearly spin-free diamond lattice provides ultrasensitive magnetic field detection[7]. The disadvantages of using a single NV for magnetic imaging are limited magnetic sensitivity, restricted vector reconstruction and the inherently slow point-scanning data acquisition. Here, we demonstrate an enhanced magnetic imaging technique with an ensemble of NV centers acting as magnetic field sensors combined with rapid CCD camera-based image detection of a large field of view. The technique offers full reconstruction of the magnetic field vector, improved temporal resolution and enhanced field sensitivity under ambient conditions.

## II. Principle and setup

The NV center in diamond consists of a substitutional nitrogen atom adjacent to a carbon vacancy (Fig. 1a). Its two unpaired electrons form a spin triplet in the electronic ground state ($^3A$), while the degenerated $m_s=\pm1$ sublevels are separated from $m_s=0$ by 2.87 GHz (Fig. 1b). The $m_s=0$ sublevel exhibits a high fluorescence rate when optically pumped. In contrast, when the center is excited in the $m_s=\pm1$ levels, the NV exhibit a higher probability to cross over to the non-radiative singlet state ($^1A$) followed by a subsequent relaxation into $m_s=0$. As a result, the spin state can be optically read out with a fluorescent contrast of ~30% ($m_s=0$ 'bright', $m_s=\pm1$ 'dark').[11] When an external magnetic field is applied, the degeneracy of the spin sublevels $m_s=\pm1$ is lifted via Zeeman splitting. This causes the resonance lines to split depending on the applied magnetic field magnitude and its direction. This dependency is used in the present work for vector magnetometry as the resonant spin transitions can be probed by sweeping the microwave frequency resulting in characteristic dips in the ODMR spectrum (Fig. 2a).

Instead of using a single NV center as field sensor we employed ion implantation to create a homogenous layer of negatively charged NV centers into an ultrapure [100] type IIa diamond (Element Six). The ensemble NV sensor offers a higher magnetic sensitivity which scales with $\sqrt{N}$ due to the amplified fluorescence signal from N sensing spins. Another advantage is the improved vector reconstruction since the diamond lattice imposes four distinct tetrahedral NV orientations (Fig. 2b). The magnetic field projections along each of these axes are measured as a single composite spectrum and an algorithm developed in this study is used to reconstruct the full magnetic field vector. The magnitude $B$ and orientation ($\theta_B$, $\varphi_B$) of the external magnetic field was calculated by analyzing the ODMR spectra of each pixel individually based on an unconstrained least-square algorithm using custom written Matlab codes (Mathworks). First, we precisely determined the center frequencies of the experimental resonance transitions in the ODMR spectra ($\nu_{\text{ODMR-Fit}}$). Secondly, we solved the eigenvalues of the diagonalized Hamiltonian $H_i$ for each NV axis using random initial input parameters ($B_0$, $\theta_{B,0}$, $\varphi_{B,0}$) which allows the calculation of all resonant transitions for a given external field[12]:

$$H_i = D \cdot \hat{d}_i + g_e \mu_B B \left( sin\theta_B \cdot cos\varphi_B \cdot \hat{S}_x + sin\theta_B \cdot sin\varphi_B \cdot \hat{S}_y + cos\varphi_B \cdot \hat{S}_z \right) , \qquad [1]$$

where $\hat{d}_i$ is the fine structure tensor for each of the four NV orientations, $\hat{S}_x, \hat{S}_y, \hat{S}_z$ are the electron spin matrices, $D$ is the zero field splitting between the spin sublevels, $g_e$ the electron g-factor and $\mu_B$ is the Bohr magneton. Then, the field parameters ($B$, $\theta_B$, $\varphi_B$) are iteratively changed and an error criterion between calculated and experimentally derived resonant transitions is computed. The final magnetic field vector is given by the solution with the converged minimal error. However, C$_{3v}$ symmetry of the NV center imposes a set of four possible indistinguishable vectors if the field is aligned to one NV axis while maximal 24 possible vector orientations exist if the field is not aligned to any NV axis[12]. From this set the correct orientation can be recovered by applying a tiny bias field along one preferred axis and monitoring corresponding perturbation on the others.  Alternatively, having an *a priori* knowledge of the magnetic field direction, the NV sensor can be placed in such a manner that the vector outcome would be distinct.

In order to demonstrate full vector reconstruction of our widefield magnetometer, we fabricated gold micro-wires on top of the diamond and imaged the magnetic field distribution created by 75 mA DC currents through the wires (see Fig. 3). Optical excitation is achieved using a 532 nm laser (Verdi, Coherent) set to widefield illumination by focusing the laser onto the back-focal plane of the objective  (60x1.49, Olympus). The two dimensional fluorescence is projected onto a 512x512 pixel CCD camera (Cascade 512B, Roper Scientific) detecting the entire image at once with an integration time of only 100 ms. Our optics yield an effective pixel size of ~100 nm with a field of view of ~60x60 μm. The camera is synchronized to the microwave source (SMIQ 03B, Rohde&Schwarz) which is swept from 2.7 to 3 GHz in steps of 500. Thus, we acquire an image stack where each image represents the response of the NV sensor at a given microwave frequency kHz. This multiplexed data acquisition of a large field of view is a substantial improvement compared to slow point-scanning methods since each pixel contains the spatial information about the external magnetic field.

## III. Results

In a first step we created the NV sensor by homogenously implanting $^{15}N^+$ ions with 4 keV per atom into the diamond. The implanted nitrogen atoms have a mean depth of 6.7±2.8 nm (Fig. 4a) and are subsequently converted to NV centers by annealing the sample at 800°C under high vacuum. This thin layer of magnetic sensing NVs permits precise detection of magnetic fields located in very close proximity to the actual NV sensor. Higher NV densities have several advantages: Shorter integration times due to higher signal-to-noise ratio and enhanced sensitivity to magnetic fields. However, the nitrogen content in diamond increases with the implantation dosage due to the imperfect conversion efficiency from implanted nitrogen to NV. Since nitrogen is a paramagnetic impurity capable of dipolar coupling to the NV, elevated nitrogen content leads to dephasing and ultimately to reduced sensitivity. In order to assess the effect of higher implantation dosages on the magnetometers performance, we varied the implantation density and analyzed the line width and signal of the ODMR spectra in zero-field. As shown in Fig. 4b, a nitrogen induced broadening of the line width was observed for higher implantation dosages. On the other hand, assuming that the magnetometer achieved the desired sensitivity, higher NV densities facilitate substantially faster ODMR readout due to the elevated fluorescence signal. Therefore, depending on the experimental requirements, a compromise is required between sensitivity and acquisition speed.

For the following experiments we used an intermediate nitrogen implantation dosage with ~1000 NV centers per μm². This ensemble density facilitates a sensitive detection with an integration time per 512x512 pixel frame of only 100 ms.

In order to show that this homogenously implanted NV sensor can be used as an efficient vector magnetometer, we optically detected the resonance transitions of the NV array sensing the magnetic field created by DC current carrying micro-wires. Subsequently, each pixel was subjected to an iterative reconstruction of the magnetic field. The experimentally extracted magnetic field amplitude agrees well with the numerical simulation (Fig. 5a). In the middle section the fields of each wire are in opposite direction and counteract each other. The reason that both fields do not cancel out completely is due to the fact that the actual NV sensors are about 2 μm below the center of the micro-wires. Thus, the $B_x$ components perpendicular to the wire are additive. The iterative algorithm also reconstructs the orientation of the external magnetic field. For demonstration purposes, we deliberately placed the micro-wires along two NV projections, hence NV1 and NV3 are parallel to the magnetic field as illustrated in Fig. 2b. In principle, it is not possible to distinguish the directionality of the magnetic field without applying a second known bias field. However, in our experimental setup the field is aligned along NV1 corresponding to $\varphi_B$=0° for all pixels as the polarity of the current is the same for both wires. The unconstrained iterative algorithm detects the expected Gaussian distribution along this NV axis for all pixels with an initial guess $\varphi_{init}$=0°. The circular orientation of the magnetic field around each wire leads to a distance dependent $\theta_B$. The exact and numerically calculated course of $\theta_B$ is illustrated by the black curve in Fig. 5b. The superposition with the blue curve is due to the crystal symmetry of the two horizontally aligned NV projections NV1 and NV3 with a vertical symmetry along $\theta_B$=90°. Apart from slight deviations, it is evident that the experimentally obtained vertical angles follow the predicted values. Thus, it is possible to regain the correct magnitude as well as the orientation of the external magnetic field. Though this iterative process is limited in terms of full vector reconstruction due to symmetry effects in diamond, it is possible to recover the true magnetic field vector. For our experimental setup all $\theta$ angles can be projected to the correct NV axis according to the right hand rule (see black line in Fig. 5b). The resulting pixel-wise calculated magnetic vector field is illustrated in Fig. 5c.

## IV. Summary and Outlook

To summarize, we implemented the first imaging vector magnetometer using an array of NV spins combined with a fast camera readout. The minimum detectable field scales with $\sigma/\sqrt{N}$, where $\sigma$ is the accuracy of fitting the resonant transitions in the ODMR spectrum and N is the number of photons contributing to the informative ODMR contrast. With typical values of $\sigma$=420 kHz and N=91,000 s$^{-1}$ our ensemble magnetometer achieves an experimental magnetic sensitivity of 20 nT/$\sqrt{\text{Hz}}$. This high magnetic sensitivity together with a spatial resolution of 250 nm and functionality under ambient conditions renders our widefield-NV-magnetometer a prime candidate for a variety of applications in science and technology. Imaging magnetic signatures of nanoscopic objects in a snapshot type imaging offers the possibility to study the dynamics with high temporal resolution. One practical application for such sensitive magnetic field detection is to perform MRI at the microscopic scale or to probe ionic fluctuations across single membrane channels.[13] Multiplexed magnetic imaging of ion-flows enables a label-free approach in imaging cellular activity with optical resolution. Considerable potential for further sensitivity enhancements resides in coherent spin manipulation techniques exploiting the longer phase memory times and enhanced readout fidelities.[7, 14-16]

**Acknowledgement:**

This work was supported by EU "Miloc", DFG "Spin microscope" and Landesstiftung "Internationale Spitzenforschung".

**Figures:**

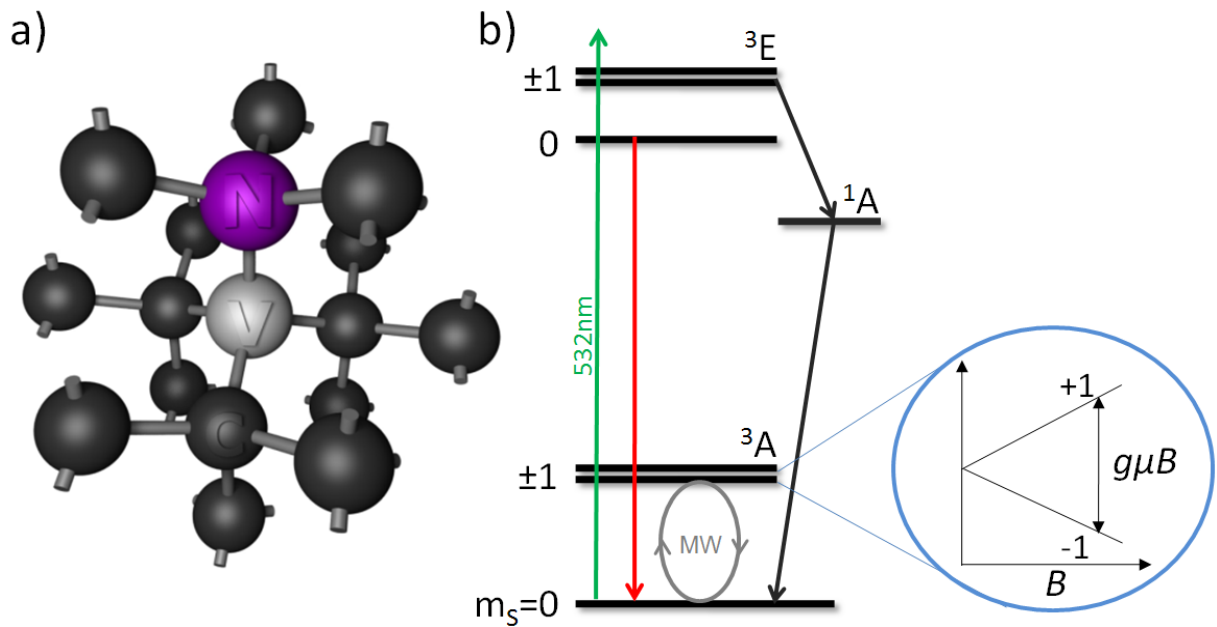

**Fig. 1.** (a) The NV center in diamond consists of a substitutional nitrogen atom (purple) associated with a vacancy (light grey). (b) Energy level scheme of the NV. The electron spin resonance transition between $m_s=0$ and $m_s=\pm1$ is changed via Zeeman splitting by an external magnetic field $B$ which lifts the degeneracy of $m_s=\pm1$ (blue circled inset). Probing the resonant spin transitions is achieved via application of microwave fields (MW) enabling the detection of the external magnetic field.

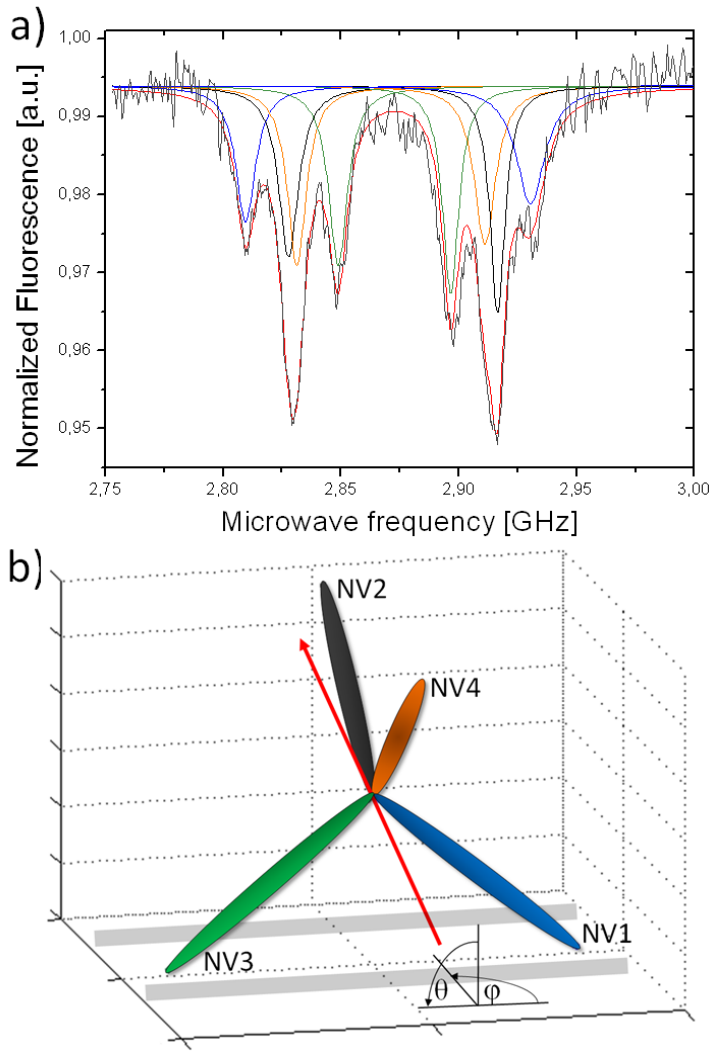

**Fig. 2.** (a) Characteristic ensemble ODMR spectrum from a single pixel. The fitted ODMR spectrum reveals the precise position of the four individual pairs of resonant transitions of each NV axis (NV1-4 color coded as shown in Fig. 2b). For this ODMR spectrum the external magnetic field can be estimated to be $B$=2.7 mT, $\theta_B$=17.7° and $\varphi_B$=6.8° (red arrow in Fig. 2b). (b) Orientation of the four crystallographic NV axes in a [100] diamond crystal. The orientation of the two micro-wires (light grey stripes) are aligned along the NV1-NV3 axis.

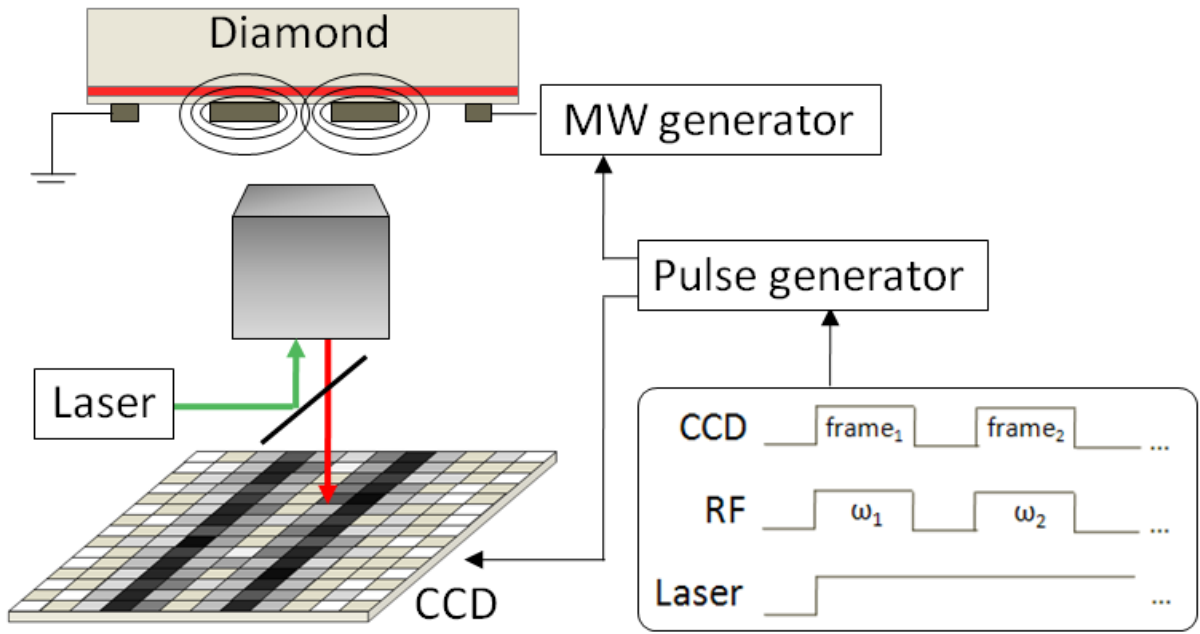

**Fig. 3** Widefield magnetic imaging setup. The shallow implanted NV centers (red layer) in diamond sense the magnetic field created by DC currents passing through the two inner wires with the same polarity (grey rectangles with black ellipses representing the magnetic field lines). Optical readout is achieved by continuously exciting the NV sensor with green laser light set to wide-field illumination, sweeping the microwaves (MW) and simultaneously detecting the entire field of view using a CCD camera.

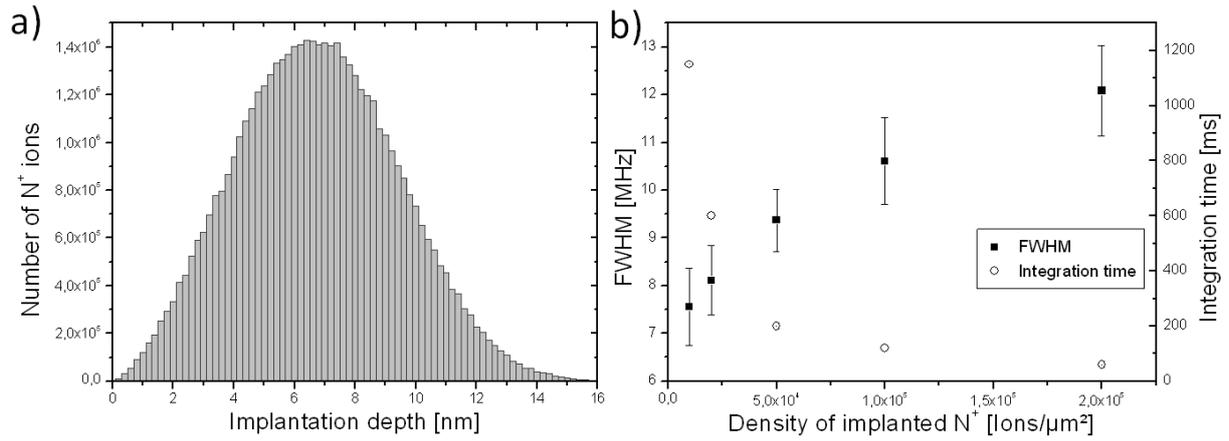

**Fig. 4.** (a) Implantation depth profile of the implanted nitrogen ions. The data is based on a Monte-Carlo simulation using the SRIM package[17]. (b) Fitted line width of the resonance peak without an external magnetic field (black rectangles, error bar represents standard deviation). Though the FWHM increased with higher NV density, significantly shorter exposure times are possible for CCD integration (dotted circles). Exposure times were set to match the same total photon counts as the highest implantation dosage.

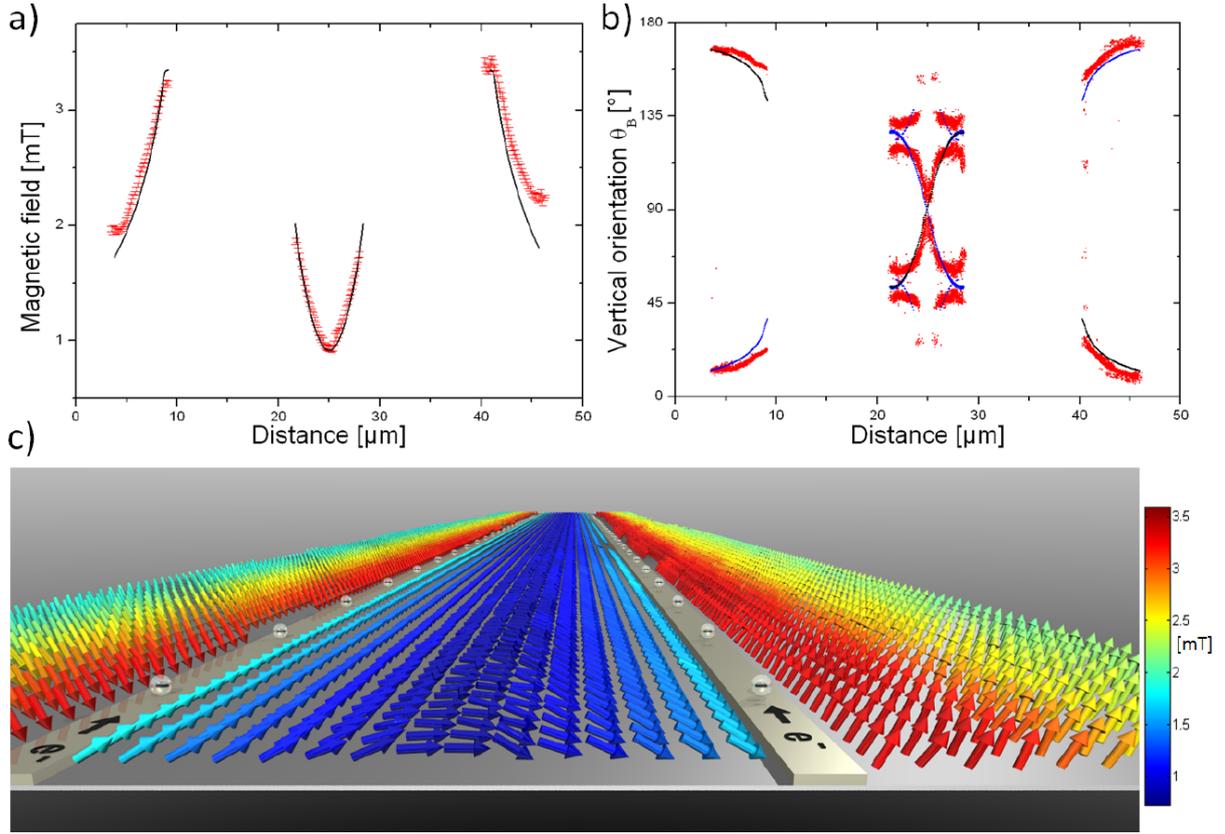

**Fig. 5.** (a) Cross-Section of the measured magnetic field averaged along the entire wire length. The magnetic field was numerically simulated (black lines) and iteratively calculated (red crosses, with standard deviation). The two wires are located at 10-20 and 30-40 μm, respectively. (b) Cross section of the simulated (black curve) and experimental evolution of $\theta_B$ (red dots). The exact solution is superimposed by a second set of possible outcomes (blue line) due to the second NV projection which is also horizontally aligned to the magnetic field. For the simulation we assumed $\varphi$ to be aligned to the NV projection ($\varphi_B=0\pm10°$) and analogously plotted the corresponding measured $\theta_B$ solutions. (c) Pixel-wise measured magnetic field ($B$, $\theta_B$, $\varphi_B$) produced by two current carrying wires. Data points perpendicular to the wires were binned by a factor of 3 and wire dimensions were downsized for representational purposes only. The plotted data was extracted from results previously shown, but $\theta_B$ angles were projected according to the right-hand rule for both wires (see black line in b).